\documentclass[aps,prl,reprint,superscriptaddress,nofootinbib,floatfix]{revtex4-2}

\usepackage{amsmath,amssymb,bm}
\usepackage{graphicx}
\usepackage{xcolor}
\usepackage[colorlinks=true,citecolor=blue,linkcolor=blue,urlcolor=blue]{hyperref}
\usepackage{pstricks, xcolor}
\usepackage{orcidlink}

\newcommand{\placeholderfig}[2]{%
  \begin{center}
  \fbox{\begin{minipage}[c][0.18\textheight][c]{0.95\linewidth}
  \centering\small Placeholder for figure file \texttt{\detokenize{#1}}.\\[0.4em]
  #2
  \end{minipage}}
  \end{center}}

\newcommand{\includewithplaceholder}[2]{%
  \IfFileExists{#1}{\includegraphics[width=\linewidth]{#1}}{\placeholderfig{#1}{#2}}}
\newcommand{\dd}{\mathrm{d}}  

\begin{document}

\title{Orthogonality Edges in Strong-Coupling Quantum Work Statistics}

\author{Atta ur Rahman~\!\!\orcidlink{0000-0001-7058-5671}}
\affiliation{School of Physical Sciences, University of Chinese Academy of Sciences, Yuquan Road 19A, Beijing 100049, China}

\author{Muhammad Noman~\!\!\orcidlink{0009-0003-6136-0114}} 
\affiliation{Institute of Fundamental Physics and Quantum Technology and School of Physical Science and Technology, Ningbo University, Ningbo 315211, China}

\author{S. M. Zangi~\!\!\orcidlink{0000-0002-4601-4681}} 
\affiliation{Department of Physics, University of Sargodha, Sargodha 40100, Pakistan}

\author{Saeed Haddadi~\!\!\orcidlink{0000-0002-1596-0763}}\email{haddadi@ipm.ir}
\address{School of Particles and Accelerators, Institute for Research in Fundamental Sciences (IPM), P.O. Box 19395-5531, Tehran, Iran}

\date{\today}
\begin{abstract}
Strong coupling to a reservoir can do more than shift, broaden, or dress the work peaks of a driven quantum system. When the reservoir is infrared singular, a sudden change of a local control parameter can alter the boundary condition seen by infinitely many low-energy modes, converting a quasiparticle-like threshold line into a many-body edge. We demonstrate this mechanism for the inclusive work distribution of the biased spin-boson model under a sudden bias inversion. In the independent-boson limit, the problem is exactly solvable and gives a sharp infrared classification: a super-Ohmic bath can retain a finite elastic threshold weight, whereas Ohmic and sub-Ohmic baths extinguish the elastic line through boundary orthogonality. At the Ohmic fixed point, the same exponent controls both the vanishing elastic residue and the low-work continuum,
$Z_{\rm el}\sim \omega_{\rm IR}^{\theta}$ and
$P_{\rm cont}(\Omega)\sim \Omega^{\theta-1}$, with $\theta=2\alpha$.
We then ask how this edge is resolved away from the static-boundary limit. Using displaced-basis exact diagonalization of logarithmically discretized baths, we find that finite tunnelling leaves an edge-like continuum over the accessible energy window, while separating two operational diagnostics of the threshold: the cumulative-continuum exponent extracted from $z$-interleaved spectra lies above the elastic-overlap exponent extracted from $z$-averaged overlaps, $\theta_C>\theta_Z$. We interpret this separation as a finite-energy crossover away from the static-boundary fixed point, not as evidence for a new asymptotic fixed point. The separation survives fitting-window variation, oscillator-cutoff checks, spectrum-size checks, and leave-one-$z$-out tests, while time-domain characteristic functions provide a compatible but non-decisive diagnostic. Finally, the same threshold edge controls the sampling cost of Jarzynski-type exponential averages, making rare low-work events increasingly important at low temperature. These results identify quantum work statistics as a direct probe of boundary orthogonality in strongly coupled open quantum systems.
\end{abstract}

\maketitle

\paragraph*{Introduction.---}
Work in quantum mechanics is not represented by a single Hermitian operator. In the standard two-projective-measurement construction, it is inferred from an energy measurement before a protocol and a second energy measurement after it~\cite{Talkner2007,Esposito2009}. For weakly coupled open systems, this definition is often reduced to transitions of the driven system Hamiltonian alone. At strong coupling, however, that reduction is generally not justified. The measured energy is the energy of the full system--reservoir composite, including interaction energy, bath rearrangement energy, and correlations generated by the driving protocol~\cite{Campisi2009,TalknerCampisi2009,Jarzynski2004,TalknerHanggi2020}. The work distribution is therefore a many-body spectral function, rather than a simple record of transitions between a few reduced-system levels. This is one reason strong-coupling stochastic thermodynamics remains both conceptually subtle and physically rich~\cite{Grabert1984,Ford1985,Subasi2012,Thingna2012,Ingold2009,Strasberg2016,Perarnau2018,Strasberg2019,Rivas2020,CresserAnders2021,Trushechkin2022}.
A useful way to organize strong-coupling effects is to ask whether the reservoir primarily renormalizes the system or whether it changes the structure of the spectrum being measured. A recent paper on work statistics in a strongly coupled spin-boson model with a super-Ohmic reservoir showed that a polaron frame can absorb the dominant dressing cloud nonperturbatively, leaving a residual interaction amenable to a master-equation treatment~\cite{Diba2024}. In that regime, the work distribution displays familiar strong-coupling signatures: shifted gaps, broadened structures, and phonon-assisted features. These effects are important, but they still belong to a quasiparticle-like picture in which identifiable threshold events can survive. The reason is infrared: for a super-Ohmic bath, the static displacement cloud associated with changing the impurity configuration has a finite infrared norm.

Ohmic and sub-Ohmic reservoirs are qualitatively different. A local bias quench then changes the boundary condition seen by a continuum of low-frequency modes. The associated many-body displacement clouds are not merely different finite coherent clouds; in the infrared limit, they become orthogonal. This is the bosonic analogue of Anderson's orthogonality catastrophe and of the threshold singularities familiar from x-ray edge physics~\cite{Anderson1967,Mahan1967,Nozieres1969,Leggett1987}. A work measurement offers a particularly direct probe of this physics. The first energy measurement prepares the ground state of one boundary condition, while the second decomposes that state in the eigenbasis of another. If the two boundary conditions are orthogonal, the nominal elastic work event cannot retain finite spectral weight.
This changes the central question. The issue is not whether strong coupling shifts a line, broadens a line, or modifies sideband weights. The sharper question is what replaces the elastic threshold line when the reservoir cloud required by the final Hamiltonian is orthogonal to the one prepared initially. We show that the replacement is an orthogonality edge: the threshold delta function is extinguished, and its weight is redistributed into a continuum at arbitrarily small excess work. The edge is not a small correction to weak-coupling thermodynamics. It is a boundary spectral singularity written directly into the full counting statistics of work.

This Letter develops the above statement in two steps. First, we solve the independent-boson limit exactly. This fixed point gives a transparent infrared classification. Super-Ohmic reservoirs can retain a finite elastic residue. Ohmic reservoirs show algebraic extinction of the residue. Sub-Ohmic reservoirs suppress it even more strongly. In the Ohmic case, the continuum edge is also exactly known, and the same exponent controls both the loss of the elastic line and the growth of the threshold continuum.
Second, we turn on finite tunnelling. Second, we turn on finite tunnelling. The spin is then no longer a strictly static boundary condition, so the exact fixed-point identity between the elastic and continuum exponents should not be assumed to survive unchanged over finite resolved energy windows. We study this regime by exact diagonalization in a spin-conditioned displaced oscillator basis. The resulting data support a robust but deliberately limited conclusion: finite tunnelling produces a finite-energy crossover in which the continuum remains edge-like over the accessible window, while the exponent inferred from the cumulative continuum is shifted above the exponent inferred from the elastic overlap. This wording is important. We do not claim that finite-bath exact diagonalization proves a new thermodynamic fixed point. Instead, we identify a controlled finite-energy signature of boundary-flow physics in a work distribution that is tied directly to the operational definition of quantum work. The evidence comes from two independent extractions: elastic overlaps are geometrically averaged over shifted logarithmic meshes before fitting, while transition spectra are interleaved before constructing the cumulative continuum. The exponent separation survives the main numerical stress tests available within the calculation. Work statistics therefore provide a sensitive diagnostic of infrared many-body rearrangement, beyond reduced-system energetics and low-order work moments.

\paragraph*{Model and work distribution.---}
Let us consider the biased spin-boson Hamiltonian
\begin{equation}
H(\epsilon)=\frac{\epsilon}{2}\sigma_z+\frac{\Delta}{2}\sigma_x+
\sum_k\omega_k b_k^\dagger b_k+
\sigma_z\sum_k g_k\bigl(b_k^\dagger+b_k\bigr),
\label{eq:H}
\end{equation}
where $\epsilon$ is a controllable bias, $\Delta$ is the tunnelling amplitude, and $b_k$ annihilates a reservoir oscillator of frequency $\omega_k$. The reservoir is characterized in the continuum by
\begin{equation}
J_s(\omega)=\alpha\,\omega_c^{1-s}\omega^s e^{-\omega/\omega_c}.
\label{eq:Js}
\end{equation}
The exponent $s$ fixes the infrared class: $s>1$ is super-Ohmic, $s=1$ is Ohmic, and $s<1$ is sub-Ohmic \cite{Leggett1987,Nazir2009,McCutcheonNazir2011,NazirMcCutcheon2016}. The protocol is a sudden inversion of the bias,
\begin{equation}
\epsilon_i=-\epsilon_0,
\qquad
\epsilon_f=+\epsilon_0,
\label{eq:biasquench}
\end{equation}
starting from the many-body ground state $|\Psi_i^0\rangle$ of $H_i\equiv H(\epsilon_i)$. If $|\Psi_f^m\rangle$ are eigenstates of $H_f\equiv H(\epsilon_f)$ with eigenvalues $E_f^m$, the zero-temperature inclusive work distribution is
\begin{equation}
P(W)=\sum_m \bigl|\langle \Psi_f^m|\Psi_i^0\rangle\bigr|^2
\delta\!\left[W-(E_f^m-E_i^0)\right].
\label{eq:TPM}
\end{equation}
This is the full composite work distribution. It includes the energy required to rearrange the bath and the interaction cloud; it is not a reduced-system transition probability.

It is useful to isolate the threshold. We define
\begin{equation}
W_{\rm th}=E_f^0-E_i^0,
\qquad
\Omega=W-W_{\rm th},
\label{eq:threshold}
\end{equation}
where $\Omega$ is the excess work above the final ground-state threshold. The distribution may then be written as
\begin{equation}
P(W)=Z_{\rm el}\,\delta(\Omega)+P_{\rm cont}(\Omega),
\label{eq:split}
\end{equation}
with an elastic threshold weight
\begin{equation}
Z_{\rm el}=\bigl|\langle \Psi_f^0|\Psi_i^0\rangle\bigr|^2.
\label{eq:Zel}
\end{equation}
A finite $Z_{\rm el}$ means that the quench can connect the two many-body ground states without creating low-energy reservoir excitations. Orthogonality means that this event disappears in the infrared limit. Since probability is conserved, the missing elastic weight must appear in transitions with arbitrarily small but nonzero $\Omega$. The threshold decomposition in Eq.~\eqref{eq:split} is therefore the natural language for distinguishing strong-coupling dressing from an actual change in spectral structure.

\paragraph*{Exact orthogonality edge.---}
The independent-boson limit, $\Delta=0$, gives an exact baseline. The spin sectors are then static, and changing the bias selects between two oscillator displacement patterns. For opposite boundary conditions, the elastic cloud overlap is
\begin{equation}
Z_{\rm el}(\omega_{\rm IR})=
\exp\!\left[-2\int_{\omega_{\rm IR}}^{\infty}\dd \omega\,\frac{J_s(\omega)}{\omega^2}\right],
\label{eq:Zgeneral}
\end{equation}
where $\omega_{\rm IR}$ is an infrared cutoff. Substituting Eq.~\eqref{eq:Js} gives
\begin{equation}
\ln Z_{\rm el}=-2\alpha\,\Gamma\!\left(s-1,\omega_{\rm IR}/\omega_c\right),
\label{eq:Zgamma}
\end{equation}
with $\Gamma(a,x)$ the upper incomplete gamma function. Hence,
\begin{align}
\label{eq:Zlimits}
s>1 &: \quad Z_{\rm el}\to Z_0>0, \nonumber\\
s=1 &: \quad Z_{\rm el}\sim(\omega_{\rm IR}/\omega_c)^\theta,
\qquad \theta=2\alpha, \\
s<1 &: \quad Z_{\rm el}\to0 \quad  \text{faster than any Ohmic power law}.\nonumber
\end{align}
The corresponding running exponent,
\begin{equation}
\theta_{\rm eff}(\omega_{\rm IR})\equiv
\frac{\dd \ln Z_{\rm el}}{\dd \ln\omega_{\rm IR}}
=2\alpha\left(\frac{\omega_{\rm IR}}{\omega_c}\right)^{s-1}
 e^{-\omega_{\rm IR}/\omega_c},
\label{eq:thetaeff}
\end{equation}
flows to zero for super-Ohmic baths, to $2\alpha$ for Ohmic baths, and to infinity for sub-Ohmic baths. Thus, the fate of the threshold line is governed by the infrared class of the reservoir. It is not decided simply by whether the coupling is weak or strong on a microscopic scale.

For the Ohmic fixed point, the continuum is also exactly determined. With $x=\Omega/\omega_c$,
\begin{equation}
\omega_c P_{\rm cont}(\Omega)=\frac{x^{\theta-1}e^{-x}}{\Gamma(\theta)},
\qquad \theta=2\alpha,
\label{eq:edge}
\end{equation}
and therefore
\begin{equation}
P_{\rm cont}(\Omega)\sim\Omega^{\theta-1},
\qquad
C(\Omega)\equiv\int_0^\Omega \dd \Omega'\,P_{\rm cont}(\Omega')\sim\Omega^\theta.
\label{eq:edgecum}
\end{equation}
Equations \eqref{eq:Zlimits} and \eqref{eq:edgecum} expose the fixed-point logic. The vanishing of the elastic residue and the creation of the threshold continuum are two sides of the same boundary-orthogonality singularity. This exact result is also a useful guardrail for the finite-tunnelling analysis below: the numerical calculation must reproduce the fixed-point exponent when the spin acts as a static boundary condition, while finite tunnelling can be assessed as a controlled departure from that anchor.

\begin{figure*}[t]
\includegraphics[width=150mm]{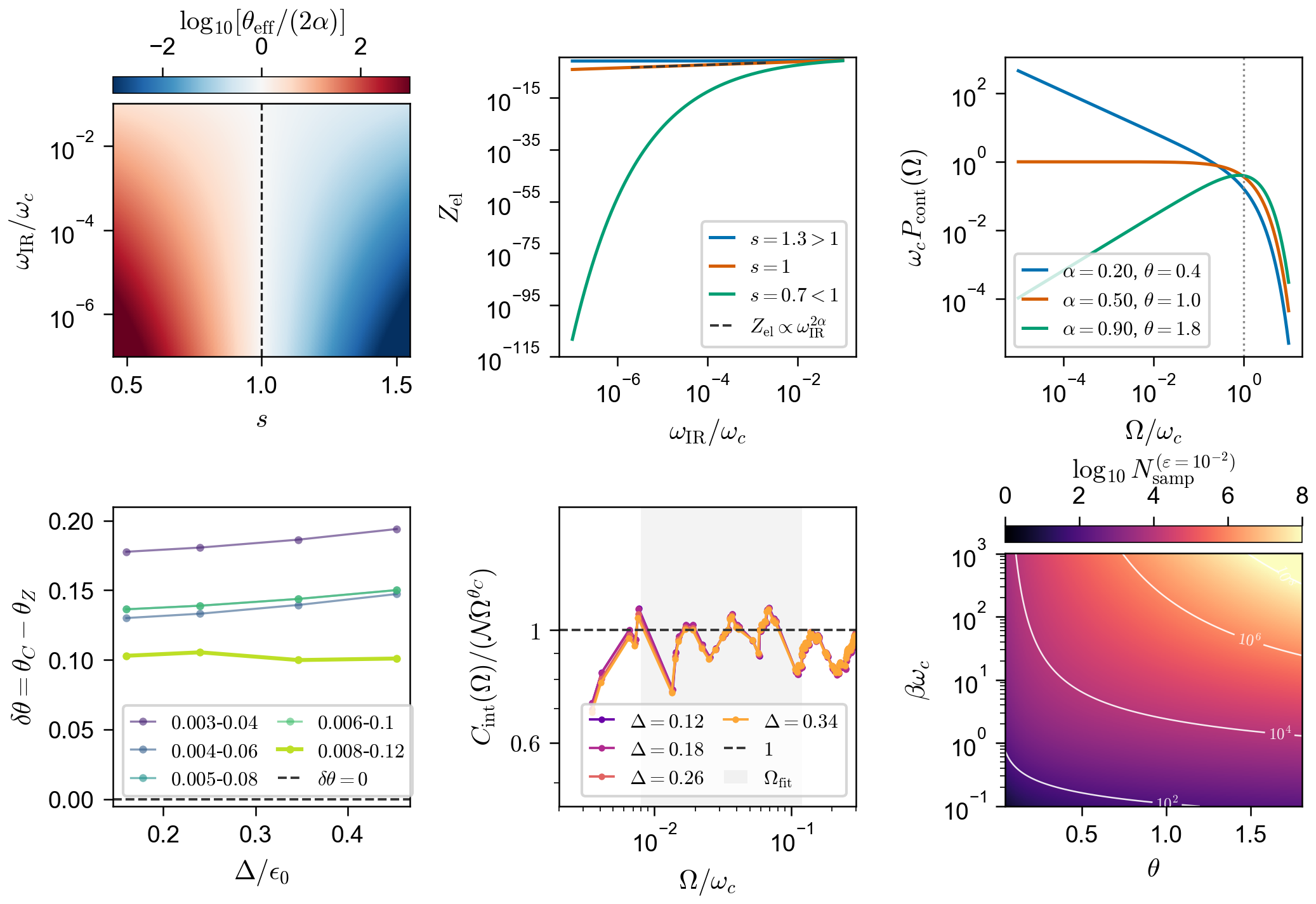}\put(-403,275){(a)}\put(-261,275){(b)}\put(-115,275){(c)}\put(-403,131){(d)}\put(-260,131){(e)}\put(-115,131){(f)}
\caption{\textbf{Orthogonality edge and finite-energy crossover in quantum work statistics.}
(a) Running infrared exponent $\theta_{\rm eff}=\dd\ln Z_{\rm el}/\dd\ln\omega_{\rm IR}$ for the spectral density $J_s(\omega)=\alpha\omega_c^{1-s}\omega^s e^{-\omega/\omega_c}$. The color scale shows $\log_{10}[\theta_{\rm eff}/(2\alpha)]$, and the dashed vertical line marks the Ohmic boundary $s=1$. The Ohmic bath is the marginal case between a finite elastic threshold line for $s>1$ and stronger infrared orthogonality for $s<1$.
(b) Elastic threshold weight $Z_{\rm el}$ as a function of the infrared cutoff $\omega_{\rm IR}/\omega_c$. The super-Ohmic bath approaches a finite residue, the Ohmic bath shows algebraic extinction $Z_{\rm el}\propto\omega_{\rm IR}^{2\alpha}$, and the sub-Ohmic bath is suppressed more rapidly.
(c) Exact Ohmic continuum edge at the independent-boson fixed point, $\omega_cP_{\rm cont}(\Omega)=x^{\theta-1}e^{-x}/\Gamma(\theta)$ with $x=\Omega/\omega_c$ and $\theta=2\alpha$. The spectral weight removed from the elastic line is redistributed into a low-work continuum.
(d) Finite-tunnelling crossover correction $\delta\theta=\theta_C-\theta_Z$ versus $\Delta/\epsilon_0$, extracted from displaced-basis exact diagonalization of logarithmically discretized Ohmic baths. Different curves correspond to different fitting windows for the cumulative-continuum exponent. Some curves nearly overlap and are visually indistinguishable within the line width because the extracted values are very close; this overlap shows that the positive finite-window shift is not sensitive to the precise fitting interval. The dashed line indicates the independent-boson identity $\delta\theta=0$.
(e) Compensated cumulative continuum $C_{\rm int}(\Omega)/(\mathcal{N}\Omega^{\theta_C})$ for several tunnelling amplitudes. The approximate plateau in the shaded fitting window shows that the low-work continuum remains edge-like over the resolved energy range.
(f) Jarzynski sampling cost associated with the Ohmic threshold edge. The color scale gives $\log_{10}N_{\rm samp}^{(\varepsilon=10^{-2})}$, showing that the number of samples required to estimate exponential work averages grows at large $\beta\omega_c$ and with increasing orthogonality exponent $\theta$.}
\label{fig:fixedpoint}
\end{figure*}
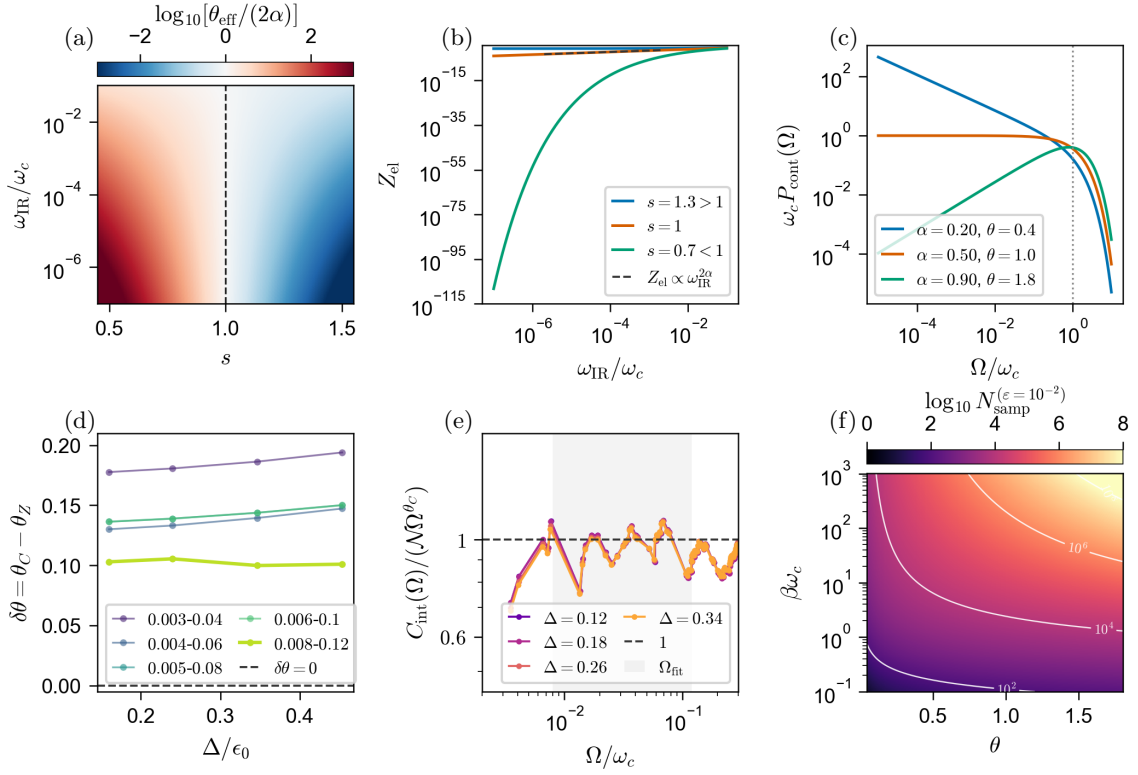

Figure~\ref{fig:fixedpoint} represents the fixed-point orthogonality edge and its finite-energy consequences in the considered configuration. 
Panels (a) to (c) show the exact independent-boson infrared classification including the running exponent and the elastic threshold weight, along with the Ohmic continuum edge. 
Panels (d) and (e) reveal how finite tunnelling can preserve an edge-like continuum while producing a positive finite-window distinction or separation between the continuum and elastic exponents. 
Panel (f) connects the same threshold singularity to the sampling cost of Jarzynski-type exponential work averages.

\paragraph*{Finite tunnelling: from a fixed point to a resolvable crossover.---}
At finite $\Delta$, the impurity is not a frozen boundary condition. Tunnelling mixes the two displaced sectors, generates an additional low-energy scale, and invalidates the simple factorization behind Eqs. (8)--(12). The elastic overlap and the continuum edge therefore become related but distinct finite-energy diagnostics. The elastic overlap compares the two many-body ground states, whereas the continuum probes the distribution of transition weight into excited states above the final threshold. At the independent-boson fixed point these diagnostics are locked by the same exponent. Away from that fixed point, we use this equality only as an infrared anchor and ask how it is resolved over the finite energy window accessible to exact diagonalization.

We investigate this regime using exact diagonalization of finite logarithmic baths in a displaced oscillator basis. The logarithmic mesh is
\begin{equation}
u_k=u_{\min}+(k+z)\Delta u,
\qquad
\omega_k=e^{u_k},
\label{eq:mesh}
\end{equation}
where $z$ shifts the grid. The discrete couplings are chosen as
\begin{equation}
g_k^2=\kappa J_s(\omega_k)\omega_k\Delta u.
\label{eq:gk}
\end{equation}
In the data shown, we use $\kappa=1/2$, so that the static Ohmic cloud benchmark gives $\theta_{\rm cloud}=2\alpha$ in our continuum convention. The basis is displaced according to the spin orientation, absorbing the large static cloud into the local oscillator basis rather than representing it by many bare Fock states. This choice is essential for an infrared problem because the orthogonality cloud is carried by low-frequency modes whose bare occupation can otherwise converge very slowly. Details of the basis construction, cutoffs, and benchmarking are given in the Supplemental Material.

We extract two exponents:
\begin{equation}
Z_{\rm el}(\omega_{\rm IR})\sim \omega_{\rm IR}^{\theta_Z},
\qquad
C(\Omega)\sim \Omega^{\theta_C}.
\label{eq:thetaZthetaC}
\end{equation}
The extraction procedures are intentionally different. For the elastic line, a single logarithmic mesh produces shell oscillations, so we first average $\ln Z_{\rm el}$ over several shifted meshes and then fit the geometric mean. For the continuum, we compute the transition weights for each shifted mesh and interleave the spectra before forming the cumulative distribution. The first operation stabilizes a ground-state overlap. The second fills artificial gaps in a sparse logarithmic spectrum. Both steps are needed before one can assign a meaningful finite-window exponent.

Figure~\ref{fig:finite_tunnelling_edge} gives the central finite-tunnelling evidence. 
Panel~(a) first anchors the finite-bath convention by showing that the discretized Ohmic 
cloud reproduces the expected static exponent, \(\theta_{\rm cloud}\simeq 2\alpha\). 
Panel~(b) then compares the two independently extracted finite-tunnelling diagnostics at 
a representative tunnelling amplitude. The elastic threshold weight is fitted as 
\(Z_{\rm el}^{\rm av}\sim \omega_{\rm IR}^{\theta_Z}\), while the interleaved cumulative 
continuum is fitted as \(C_{\rm int}(\Omega)\sim \Omega^{\theta_C}\). In this panel the 
common abscissa \(\xi\) denotes \(\omega_{\rm IR}\) for \(Z_{\rm el}^{\rm av}\) and 
\(\Omega\) for \(C_{\rm int}\). The fitted values show
\[
\theta_C>\theta_Z ,
\]
so finite tunnelling separates two threshold diagnostics that coincide at the independent-boson fixed point.
We interpret this as a finite-window crossover signature, not as a determination of a new asymptotic exponent.
Across the accessible Ohmic scaling window, this separation remains positive as: 

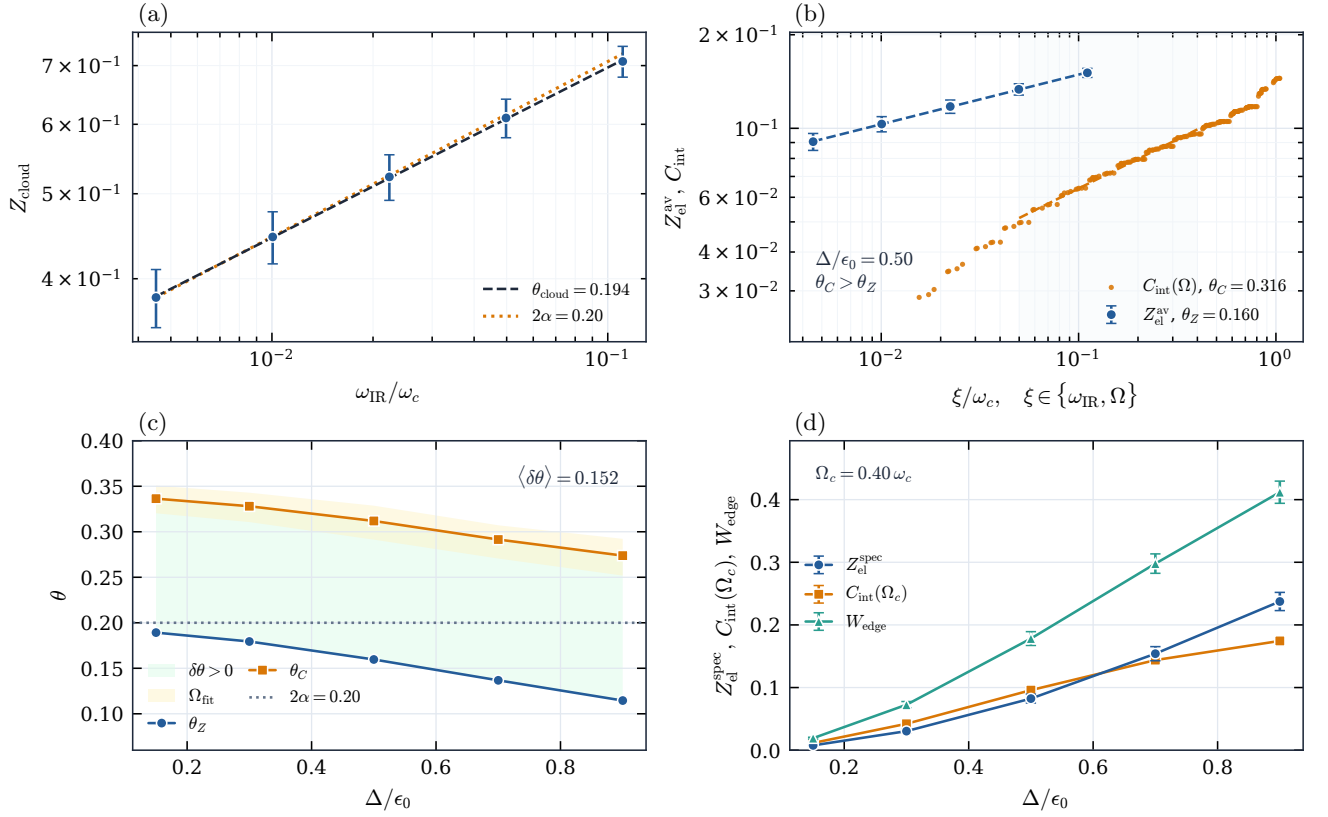
\begin{figure*}[t] \centering \includegraphics[width=0.96\textwidth]{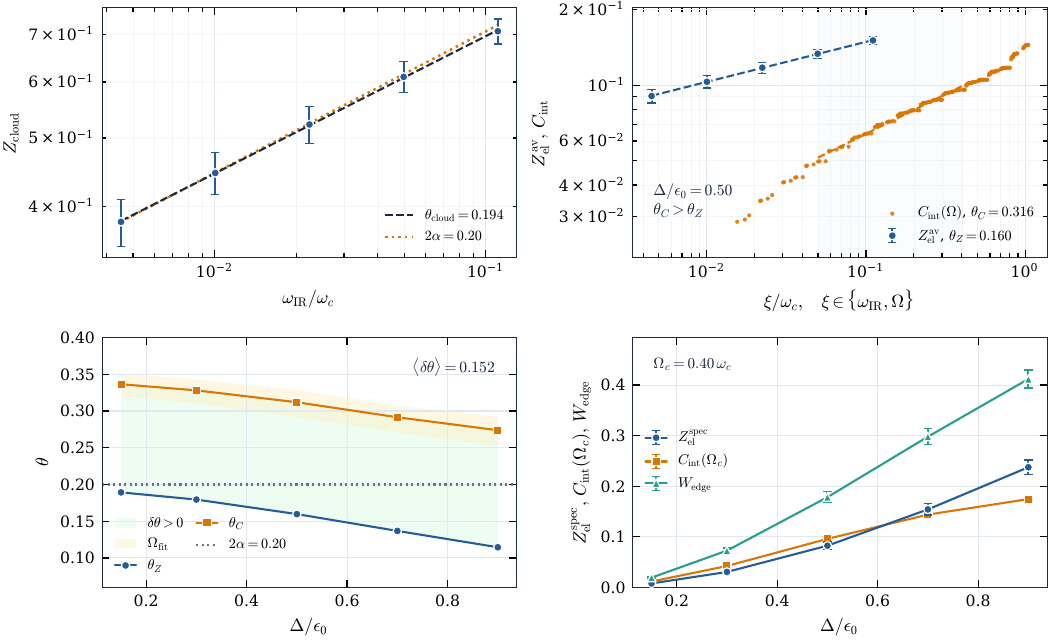}
\put(-440,300){(a)}
\put(-193,300){(b)}
\put(-440,146){(c)}
\put(-193,146){(d)} \caption{\textbf{Finite-tunnelling crossover and edge-window spectral accounting.} (a) Static Ohmic cloud benchmark for the same logarithmic discretization convention used in the finite-tunnelling calculation. The fitted exponent $\theta_{\rm cloud}$ agrees with the target value $2\alpha$, anchoring the finite-bath normalization to the independent-boson fixed point. (b) Representative finite-tunnelling extraction at $\Delta/\epsilon_0=0.50$. The elastic threshold weight is fitted as $Z_{\rm el}^{\rm av}\sim\omega_{\rm IR}^{\theta_Z}$ after geometric averaging over shifted logarithmic meshes, while the continuum weight is fitted as $C_{\rm int}(\Omega)\sim\Omega^{\theta_C}$ after interleaving the shifted spectra. The common horizontal variable $\xi$ denotes $\omega_{\rm IR}$ for $Z_{\rm el}^{\rm av}$ and $\Omega$ for $C_{\rm int}$. (c) Effective finite-window exponents versus tunnelling amplitude. The continuum exponent remains above the elastic-overlap exponent over the resolved scaling window, giving a positive finite-window separation $\delta\theta=\theta_C-\theta_Z$. The shaded region highlights $\delta\theta>0$, and the orange band indicates fitting-window variation for the continuum extraction. This panel should be read as a finite-energy crossover diagnostic, not as an asymptotic fixed-point extrapolation. (d) Edge-window spectral accounting at $\Omega_c=0.40\omega_c$. The plotted quantities are the resolved elastic contribution $Z_{\rm el}^{\rm spec}$, the low-work continuum weight $C_{\rm int}(\Omega_c)$, and their threshold-sector sum $W_{\rm edge}=Z_{\rm el}^{\rm spec}+C_{\rm int}(\Omega_c)$. This panel is a finite-window diagnostic of spectral-weight redistribution near the edge, not a full Hilbert-space sum rule.} \label{fig:finite_tunnelling_edge}\end{figure*}
{
\begin{equation}
\delta\theta\equiv \theta_C-\theta_Z>0 .
\label{eq:theta_separation}
\end{equation}
Panel~(c) displays this behavior as a function of \(\Delta/\epsilon_0\). The shaded 
region marks \(\delta\theta>0\), and the band indicates fitting-window variation of the 
continuum exponent. Panel~(d) adds a complementary edge-window accounting check: within 
a fixed low-work window \(\Omega_c\), the resolved continuum weight 
\(C_{\rm int}(\Omega_c)\) grows together with the threshold-sector weight 
\(W_{\rm edge}=Z_{\rm el}^{\rm spec}+C_{\rm int}(\Omega_c)\). This quantity is not a full 
Hilbert-space sum rule; it is a finite-window diagnostic showing that the observed edge is 
accompanied by resolved low-work spectral weight, not only by a slope fit.}

\paragraph*{Numerical safeguards and physical interpretation.---}
The most natural objections to Eq.~\eqref{eq:theta_separation} are numerical. Logarithmic baths are sparse. A finite oscillator cutoff can distort low-frequency clouds. A limited number of final states can bias a cumulative spectrum. A favorable set of $z$ shifts could accidentally create an apparent power law. The Supplemental Material is organized around these issues rather than around cosmetic convergence checks.

First, the displaced-basis implementation is benchmarked against the exactly solvable static cloud. In the Ohmic case, both the analytic discrete cloud expression and the finite displaced-basis matrix representation reproduce the expected exponent $\theta_{\rm cloud}=2\alpha$ within uncertainty. This test checks the normalization of the spectral density, the displacement convention, and the Franck-Condon matrix elements used in the tunnelling term. Second, $\theta_Z$ and $\theta_C$ are extracted from different processed objects: a geometric mean of elastic overlaps for $\theta_Z$, and an interleaved cumulative transition spectrum for $\theta_C$. The exponent separation is therefore not a single-fit artifact. Third, the positive shift survives increasing the local oscillator cutoff for infrared modes, increasing the number of retained final eigenstates in the continuum construction, varying the fitting window, and removing individual $z$ meshes in leave-one-$z$-out tests.

For each finite bath, the transition weights entering the continuum are
\begin{equation}
p_m=\bigl|\langle\Psi_f^m|\Psi_i^0\rangle\bigr|^2,
\qquad
\Omega_m=E_f^m-E_f^0,
\label{eq:transitions}
\end{equation}
and the cumulative continuum is
\begin{equation}
C(\Omega)=\sum_{0<\Omega_m<\Omega}p_m.
\label{eq:Comega_discrete}
\end{equation}
We examine both the compensated quantity $C(\Omega)/\Omega^{\theta_C}$ and the local slope
\begin{equation}
\theta_C^{\rm loc}(\Omega)=\frac{\dd \ln C(\Omega)}{\dd \ln\Omega}.
\label{eq:thetaCloc}
\end{equation}
These diagnostics show the expected finite-bath oscillations, but remain consistent with an algebraic window over the range used for the fits. In other words, the continuum behaves as a threshold edge over the resolved window, even though the fitted exponent should be understood as an effective energy-domain exponent.

We also compute a time-domain characteristic function from the same transition spectrum,
\begin{equation}
G_{\rm cont}^{\rm ED}(t)=\frac{1}{P_{\rm cont}}\sum_{m>0}p_m e^{-i\Omega_m t},
\qquad
P_{\rm cont}=\sum_{m>0}p_m.
\label{eq:Gcont}
\end{equation}
In a continuum with $P_{\rm cont}(\Omega)\sim\Omega^{\theta_C-1}$, the long-time envelope scales as $|G_{\rm cont}(t)|\sim t^{-\theta_C}$ up to finite-bandwidth corrections. The finite spectra show compatible algebraic envelopes over time windows reciprocal to the energy-domain fits. We nevertheless treat this as a diagnostic only. Logarithmically discretized finite baths produce log-periodic structures and recurrences, so the decisive evidence for the exponent separation in Eq.~\eqref{eq:theta_separation} is the energy-domain analysis of \(Z_{\rm el}\) and \(C(\Omega)\).

Taken together, these checks support a physically modest but meaningful claim. The calculation does not eliminate all finite-size limitations; no finite-bath ED calculation can. It does, however, rule out the most direct artifact explanations of the observed separation and places the result in a defensible form: finite tunnelling unlocks the fixed-point equality between elastic and continuum diagnostics over the controlled scaling window accessible to the calculation.

\paragraph*{Thermodynamic consequence: rare low-work events.---}
The same edge that reorganizes the spectrum also affects fluctuation-relation estimation. After subtracting the deterministic threshold contribution, consider the normalized Ohmic continuum and the Jarzynski-type estimator $X=e^{-\beta\Omega}$. Using Eq.~\eqref{eq:edge}, its relative variance is
\begin{equation}
\frac{{\rm Var}\,X}{\langle X\rangle^2}
=
\left[\frac{(1+\beta\omega_c)^2}{1+2\beta\omega_c}\right]^\theta-1.
\label{eq:varX}
\end{equation}
For a target relative error $\varepsilon$, the required number of independent samples scales as
\begin{equation}
N_{\rm samp}\sim \varepsilon^{-2}
\frac{{\rm Var}\,X}{\langle X\rangle^2}.
\label{eq:Nsamp}
\end{equation}
At low temperature, $\beta\omega_c\gg1$, this becomes
\begin{equation}
N_{\rm samp}\sim \varepsilon^{-2}\left(\frac{\beta\omega_c}{2}\right)^\theta.
\label{eq:Nsamp_asym}
\end{equation}
Thus, the orthogonality edge is not only a spectral signature, but also controls the statistical difficulty of estimating exponential work averages, as the average becomes increasingly sensitive to trajectories with anomalously small excess work~\cite{Jarzynski2017}. This connection is useful conceptually: the same low-work events that reveal the boundary singularity are the events that dominate fluctuation-relation sampling in the low-temperature regime.

\vspace{5mm}
\paragraph*{Conclusion.---}
We have shown that strong-coupling quantum work statistics can enter an orthogonality-edge regime. In a super-Ohmic reservoir, the bath cloud associated with a local quench has a finite infrared norm, so an elastic threshold line can survive even when strong coupling substantially renormalizes and dresses the work distribution. In Ohmic and sub-Ohmic reservoirs, the infrared structure is different. A sudden bias inversion changes the reservoir boundary condition in a way that requires infinitely many low-frequency modes to rearrange, causing the corresponding many-body clouds to become orthogonal. The elastic threshold event is then extinguished, and the lost weight appears as a low-work continuum.
At the exactly solvable Ohmic independent-boson fixed point, the elastic extinction and continuum creation are governed by one exponent, $\theta=2\alpha$. Finite tunnelling turns this fixed-point identity into a finite-energy diagnostic problem. Within the controlled ED scaling window, the continuum remains edge-like, while the cumulative-continuum exponent lies above the exponent extracted from the elastic overlap. We therefore interpret the observed separation as an energy-domain crossover away from the static-boundary limit, not as proof of a new asymptotic fixed point. Its robustness under independent extraction procedures and numerical stress tests makes it a concrete finite-window signature of boundary-flow physics in full work statistics.
The broader message is that inclusive work distributions can reveal many-body infrared physics that is invisible to reduced-system thermodynamics and to low-order energy moments. In the present model, the work distribution directly resolves the replacement of an elastic threshold line by an orthogonality edge. The same singularity also controls the rare-event cost of Jarzynski-type estimation. These features make strong-coupling work statistics a useful meeting point between nonequilibrium thermodynamics, impurity physics, and boundary critical phenomena.

In summary, our results establish quantum work statistics as a sensitive and direct probe of boundary orthogonality in strongly coupled open quantum systems. The emergence of a universal infrared edge structure, governed by the same exponent controlling Anderson orthogonality, indicates that the observed behavior is not restricted to the biased spin-boson model but should extend to a broad class of impurity problems with infrared-singular environments. This suggests a general mechanism whereby sudden parameter changes translate many-body boundary rearrangements into measurable singular features in nonequilibrium work distributions. Importantly, the predicted edge physics is expected to be accessible in current experimental platforms, including superconducting circuit QED architectures, semiconductor quantum dots, and engineered quantum impurity simulators in ultracold atomic systems, where controlled quenches and high-resolution energy or work measurements are becoming feasible. These findings open a pathway for using nonequilibrium fluctuation statistics as a practical diagnostic tool for many-body boundary criticality in quantum open systems.

\bibliography{ref}


\clearpage
\onecolumngrid
\section*{Supplemental Material: Numerical protocol and robustness of the finite-tunnelling edge}

This Supplemental Material gives the numerical details behind the finite-tunnelling exponent analysis reported in the main text. The main text makes two related claims. First, in the independent-boson limit, the Ohmic work distribution has an orthogonality edge anchored by an exactly solvable boundary fixed point. Second, when finite tunnelling is included, displaced-basis exact diagonalization resolves a robust energy-domain separation between the exponent inferred from the elastic overlap and the exponent inferred from the cumulative continuum. The purpose of this supplement is to define the extraction protocol and to document why the observed separation is not naturally explained by logarithmic shell effects, oscillator truncation, finite spectrum size, or a single favorable logarithmic mesh.

Throughout, $Z_{\rm el}$ denotes the elastic threshold weight, $C(\Omega)$ denotes the cumulative continuum weight, and the fitted exponents are defined by
\begin{equation}
Z_{\rm el}\sim\omega_{\rm IR}^{\theta_Z},
\qquad
C(\Omega)\sim\Omega^{\theta_C}.
\label{eq:supp_def}
\end{equation}
The statement $\theta_C>\theta_Z$ should be read in the same restricted sense used in the main text: it is a finite-energy scaling-window result obtained from the numerical protocol described below, not a claim that the calculation proves a new thermodynamic fixed point.

\subsection*{A. Logarithmic bath discretization}

For each infrared cutoff $\omega_{\rm IR}$ and logarithmic shift $z$, the reservoir is discretized on the mesh
\begin{equation}
u_k=u_{\min}+(k+z)\Delta u,
\qquad
\omega_k=e^{u_k},
\qquad
u_{\min}=\ln\omega_{\rm IR},
\label{eq:supp_mesh}
\end{equation}
with $k=0,\ldots,N_b-1$. The discrete couplings are chosen as
\begin{equation}
g_k^2=\kappa J(\omega_k)\omega_k\Delta u,
\label{eq:supp_gk}
\end{equation}
where $\kappa$ fixes the convention relating the continuum spectral density to the finite-bath Hamiltonian. In the data shown in the main text and in the figures below, we use $\kappa=1/2$. With this convention, the static Ohmic cloud benchmark gives $\theta_{\rm cloud}=2\alpha$.

Logarithmic discretization is well suited to a threshold problem because the edge is controlled by modes spread over many decades of energy. The same discretization also introduces a familiar complication: any one shifted logarithmic mesh is sparse and shell dependent. For that reason, no exponent in the main text is extracted from a single raw mesh. Elastic overlaps are averaged over shifts before fitting, and continuum spectra are interleaved over shifts before constructing the cumulative distribution.

\subsection*{B. Spin-conditioned displaced basis}

The Ohmic problem is dominated by low-frequency displacement clouds. A bare oscillator Fock basis is inefficient for such states because the static displacement of the lowest modes can require many bare quanta. To avoid confusing basis convergence with infrared physics, we use a spin-conditioned displaced basis,
\begin{equation}
|\sigma;\{n_k\}\rangle_{\rm dis}
=|\sigma\rangle\prod_k
D\!\left[-\sigma\frac{g_k}{\omega_k}\right]|n_k\rangle,
\label{eq:supp_displaced_basis}
\end{equation}
where $\sigma=\pm1$ is the eigenvalue of $\sigma_z$, and $D(\lambda)=\exp[\lambda(b^\dagger-b)]$ for real $\lambda$. In this basis, the dominant diagonal displacement is already included in the local basis. The tunnelling term is represented by Franck-Condon matrix elements,
\begin{equation}
S_{\{n_k\},\{m_k\}}=
\prod_k
\left\langle n_k\left|
D\!\left(2g_k/\omega_k\right)
\right|m_k\right\rangle.
\label{eq:supp_fc}
\end{equation}
The local oscillator cutoff is mode-dependent. The lowest-frequency modes, which carry the largest part of the orthogonality cloud, are assigned larger local Hilbert spaces. Higher-frequency modes use smaller cutoffs. This allocation is a practical compromise: it targets the infrared sector relevant to the edge without making the full tensor-product Hilbert space prohibitively large.

\subsection*{C. Static cloud benchmark}

Before studying finite tunnelling, the implementation is checked against the exactly solvable static cloud. For opposite boundary conditions,
\begin{equation}
Z_{\rm cloud}=
\prod_k
\left|\left\langle 0\left|D\!\left(2g_k/\omega_k\right)\right|0\right\rangle\right|^2
=
\exp\!\left[-\sum_k\left(\frac{2g_k}{\omega_k}\right)^2\right].
\label{eq:supp_cloud}
\end{equation}
For an Ohmic bath and the convention in Eq.~\eqref{eq:supp_gk}, this gives
\begin{equation}
Z_{\rm cloud}\sim\omega_{\rm IR}^{\theta_{\rm cloud}},
\qquad
\theta_{\rm cloud}=2\alpha.
\label{eq:supp_cloud_exp}
\end{equation}
This benchmark tests three ingredients at once: the normalization of the discrete spectral density, the sign and magnitude of the displacement convention, and the finite-basis representation of the Franck-Condon matrices. As shown in Fig.~\ref{fig:supp_audit}(a), both the exact discrete cloud expression and its finite displaced-basis matrix evaluation reproduce the expected Ohmic exponent within uncertainty. This does not by itself prove the finite-tunnelling result, but it verifies that the calculation reproduces the known infrared exponent in the limit where that exponent is exact.

\subsection*{D. Extraction of $\theta_Z$: average before fitting}

For finite tunnelling, the elastic overlap is evaluated for every shifted mesh and every infrared cutoff,
\begin{equation}
Z_{\rm el}^{(z)}(\omega_{\rm IR})=
\left|\langle \Psi_{f,z}^0(\omega_{\rm IR})|\Psi_{i,z}^0(\omega_{\rm IR})\rangle\right|^2.
\label{eq:supp_Zz}
\end{equation}
A single shifted logarithmic mesh can show visible shell oscillations. We therefore average the logarithm of the overlap,
\begin{equation}
\overline{\ln Z_{\rm el}}(\omega_{\rm IR})=
\frac{1}{N_z}\sum_z \ln Z_{\rm el}^{(z)}(\omega_{\rm IR}),
\qquad
Z_{\rm el}^{\rm av}=\exp\!\left[\overline{\ln Z_{\rm el}}\right],
\label{eq:supp_zavg_Z}
\end{equation}
and then fit
\begin{equation}
Z_{\rm el}^{\rm av}(\omega_{\rm IR})\sim\omega_{\rm IR}^{\theta_Z}.
\label{eq:supp_thetaZ}
\end{equation}
The uncertainty assigned to $\theta_Z$ combines fitting-window variation with a leave-one-$z$-out jackknife component. The use of the geometric mean is appropriate because the overlap is a multiplicative infrared quantity, and because logarithmic fluctuations in $Z_{\rm el}$ are more nearly additive than fluctuations in $Z_{\rm el}$ itself.

\subsection*{E. Extraction of $\theta_C$: interleave before fitting}

For the continuum edge, the transition spectrum is computed for every shifted mesh:
\begin{equation}
p_m^{(z)}=\left|\langle\Psi_{f,z}^m|\Psi_{i,z}^0\rangle\right|^2,
\qquad
\Omega_m^{(z)}=E_{f,z}^m-E_{f,z}^0.
\label{eq:supp_pm}
\end{equation}
Rather than fitting the staircase generated by a single sparse spectrum, we interleave all shifted spectra into one averaged cumulative distribution,
\begin{equation}
C_{\rm int}(\Omega)=
\frac{1}{N_z}\sum_z\sum_{0<\Omega_m^{(z)}<\Omega}p_m^{(z)}.
\label{eq:supp_Cint}
\end{equation}
The exponent is then extracted from
\begin{equation}
C_{\rm int}(\Omega)\sim\Omega^{\theta_C}.
\label{eq:supp_thetaC}
\end{equation}
Interleaving suppresses artificial gaps between logarithmic shells while preserving the low-energy ordering of spectral weight. The uncertainty assigned to $\theta_C$ combines fitting-window variation with leave-one-$z$-out jackknife variation. This protocol makes the continuum exponent independent of the elastic-overlap extraction, which is important for interpreting the separation $\theta_C-\theta_Z$.

\renewcommand{\thefigure}{S\arabic{figure}}
\setcounter{figure}{0}

\begin{figure*}[t]
\centering
\includewithplaceholder{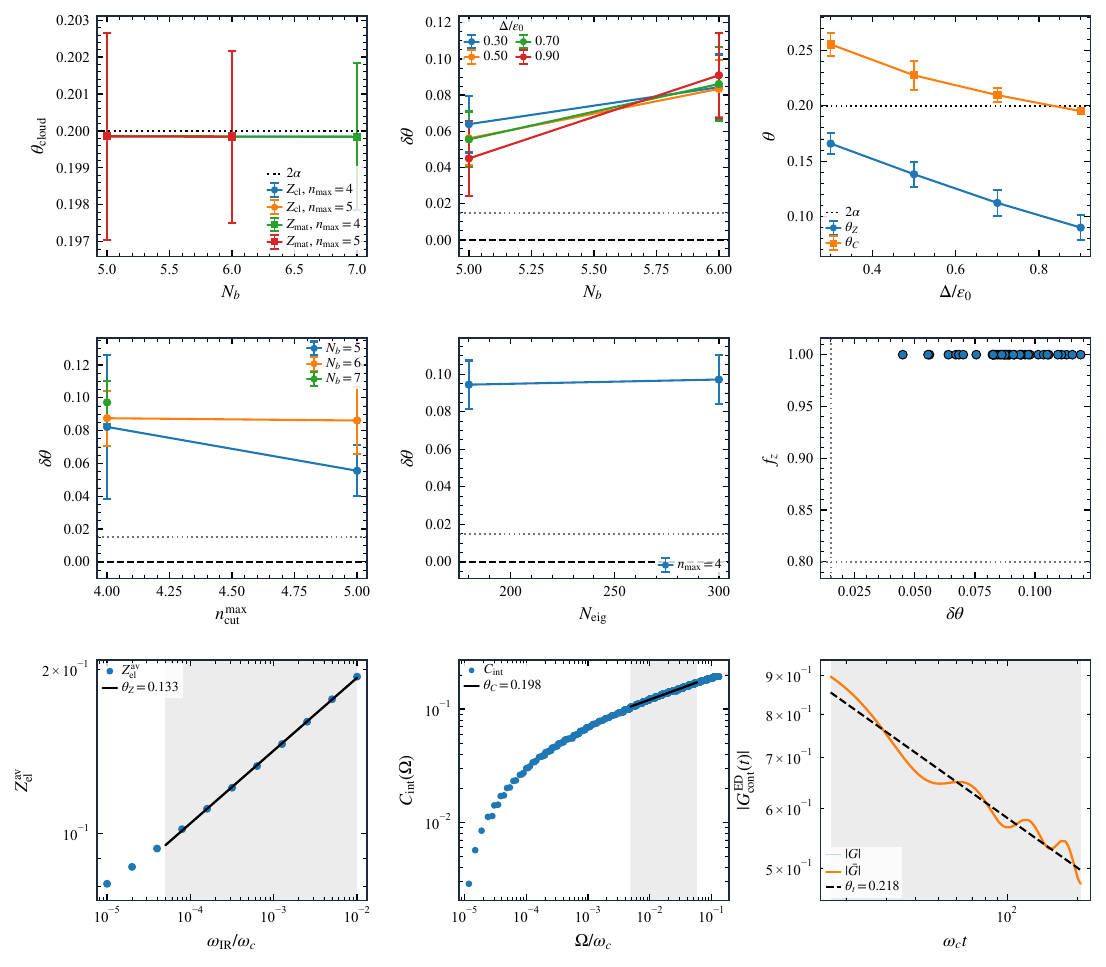}{Use this slot for the final displaced-basis ED audit figure.}
\put(-464,442){(a)}
\put(-297,442){(b)}
\put(-130,442){(c)}
\put(-464,293){(d)}
\put(-297,293){(e)}
\put(-130,293){(f)}
\put(-464,145){(g)}
\put(-297,145){(h)}
\put(-130,145){(i)}
\caption{\textbf{Numerical audit of the finite-tunnelling threshold exponents.}
(a) Static displaced-cloud benchmark. Both the exact cloud expression and the finite displaced-basis matrix representation reproduce the expected Ohmic exponent $\theta_{\rm cloud}=2\alpha$. In panel (a), the exact cloud expression and the finite displaced-basis matrix results for different oscillator cutoffs nearly overlap within the plotting resolution, confirming convergence of the static Ohmic cloud benchmark to the expected exponent $\theta_{\rm cloud}=2\alpha$.
(b) Energy-domain exponent separation $\delta\theta=\theta_C-\theta_Z$ for several tunnelling amplitudes. The dashed line marks $\delta\theta=0$, and the dotted line indicates the conservative positive-shift threshold used in the audit.
(c) Separately extracted exponents $\theta_Z$ and $\theta_C$ after $z$ averaging and spectral interleaving. The continuum exponent remains above the elastic-overlap exponent over the tested finite-tunnelling range.
(d) Local-basis convergence audit at representative tunnelling. The positive shift survives increasing the maximum oscillator cutoff used for the infrared modes.
(e) Spectrum-size audit. Increasing the number of retained final eigenstates does not remove the positive separation.
(f) Leave-one-$z$-out stability. The positive separation is stable under the removal of individual logarithmic meshes.
(g) Representative elastic-overlap fit using $Z_{\rm el}^{\rm av}=\exp[\langle\ln Z_{\rm el}\rangle_z]$.
(h) Representative $z$-interleaved cumulative continuum fit, $C_{\rm int}(\Omega)\sim\Omega^{\theta_C}$.
(i) Diagnostic time-domain envelope constructed from the same interleaved transition spectrum. This panel is used as a consistency check because logarithmic ED spectra contain finite-size recurrences.}
\label{fig:supp_audit}
\end{figure*}

\subsection*{F. Main numerical audit}

Figure~\ref{fig:supp_audit} summarizes the numerical audit. Panel (a) shows the static cloud benchmark. Panels (b) and (c) compare $\theta_Z$ and $\theta_C$ and show a positive finite-window separation,
\begin{equation}
\delta\theta=\theta_C-\theta_Z>0,
\label{eq:supp_dtheta}
\end{equation}
over the tested finite-tunnelling range. Panels (d) and (e) show that the separation is stable under increasing the maximum local oscillator cutoff and under increasing the number of final eigenstates retained in the continuum construction. Panel (f) shows leave-one-$z$-out stability. Panels (g) and (h) display representative elastic and continuum fits after the appropriate $z$ processing. Panel (i) shows the time-domain envelope constructed from the same transition spectrum. That last panel is used only as a diagnostic because finite logarithmic spectra contain recurrences.

The audit supports the limited conclusion used in the main text. Displaced-basis ED, combined with $z$-averaged elastic overlaps and $z$-interleaved transition spectra, resolves a positive energy-domain separation between the continuum and elastic exponents over the accessible Ohmic scaling window. The checks make the most direct artifact explanations unlikely, but they do not turn a finite-bath calculation into a proof of a new asymptotic fixed point.

\subsection*{G. Continuum scaling diagnostics}

\begin{figure*}[t]
\centering
\includewithplaceholder{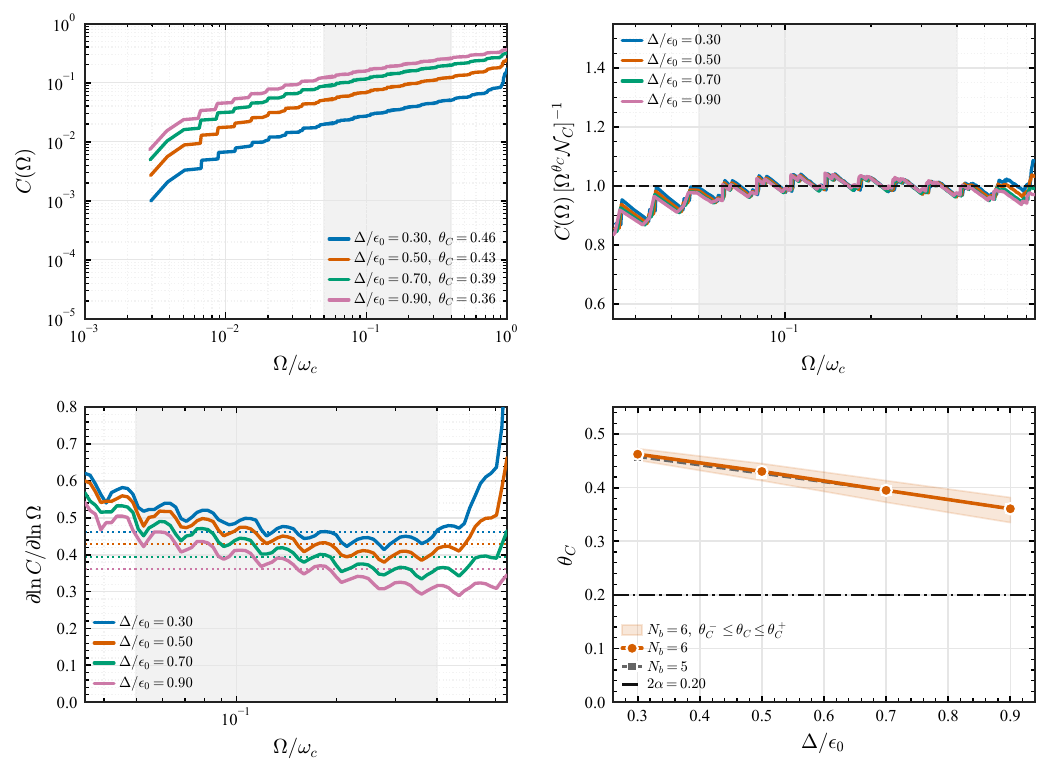}{Use this slot for the continuum-scaling audit figure.}
\put(-465,370){(a)}
\put(-210,370){(b)}
\put(-465,183){(c)}
\put(-210,183){(d)}
\caption{\textbf{Finite-tunnelling continuum scaling.}
(a) Cumulative continuum weight $C(\Omega)$ for several tunnelling amplitudes; dashed lines indicate power-law fits $C(\Omega)\sim\Omega^{\theta_C}$ over the shaded fitting window.
(b) Compensated data, $C(\Omega)/\Omega^{\theta_C}$, showing approximate plateaus over the same window.
(c) Local running exponent $\dd \ln C/\dd \ln\Omega$ with dotted lines marking the fitted values.
(d) Extracted $\theta_C$ versus $\Delta/\epsilon_0$, with the shaded band indicating fitting-window variation and the fixed-point value $2\alpha$ shown for reference.}
\label{fig:supp_continuum}
\end{figure*}

The exponent separation would be less meaningful if the continuum fit were simply a curved finite-window crossover forced into a power-law form. We therefore examine the raw cumulative continuum, the compensated cumulative continuum, and the local slope. The quantities used are
\begin{equation}
C(\Omega)=\sum_{0<\Omega_m<\Omega}p_m,
\qquad
C_{\rm comp}(\Omega)=\frac{C(\Omega)}{\Omega^{\theta_C}},
\qquad
\theta_C^{\rm loc}(\Omega)=\frac{\dd \ln C}{\dd \ln\Omega}.
\label{eq:supp_cont_checks}
\end{equation}
As shown in Fig.~\ref{fig:supp_continuum}, the compensated spectra form approximate plateaus over the same windows used for the fits, and the local slopes remain near the fitted values up to finite-bath oscillations. These diagnostics support the statement that the continuum remains threshold-edge controlled in the finite-tunnelling calculations. They also help identify the fitting window: the window is chosen where the compensated data are least curved and where the local slope is sufficiently stable before ultraviolet curvature and finite-size saturation become dominant.

\subsection*{H. Reciprocal time-domain diagnostic and fit-window stability}

The continuum characteristic function is constructed directly from the transition spectrum,
\begin{equation}
G_{\rm cont}^{\rm ED}(t)=\frac{1}{P_{\rm cont}}\sum_{m>0}p_m e^{-i\Omega_m t},
\qquad
P_{\rm cont}=\sum_{m>0}p_m .
\label{eq:supp_Gcont}
\end{equation}
In the thermodynamic continuum limit, an edge
\(P_{\rm cont}(\Omega)\sim\Omega^{\theta_C-1}\) implies an algebraic
long-time envelope \( |G_{\rm cont}(t)|\sim t^{-\theta_C}\), up to
finite-bandwidth corrections. In a finite logarithmically discretized bath,
however, \(G_{\rm cont}^{\rm ED}(t)\) also contains log-periodic oscillations,
near-zeros, and recurrences. For this reason, the time-domain signal is used
only as a reciprocal-window diagnostic, while the primary exponent is extracted
from the energy-domain cumulative continuum \(C(\Omega)\).

To test whether the energy-domain exponent is tied to a single selected
interval, we also scan the fitting bounds and extract
\begin{equation}
\theta_C(\Omega_{\min},\Omega_{\max})
\quad \text{from} \quad
C(\Omega)\sim \Omega^{\theta_C}
\label{eq:supp_window_map}
\end{equation}
for every admissible window satisfying the minimum-width and unsaturated-weight
criteria. The selected window \(\mathcal W_*\) is then compared with neighboring
windows in the \((\Omega_{\min},\Omega_{\max})\) plane. This window-stability
map is a stronger diagnostic than a single local derivative because it directly
tests whether the reported \(\theta_C\) is produced by a broad region of
acceptable fits or by one cherry-picked interval.

\begin{figure*}[t]
\centering
\includewithplaceholder{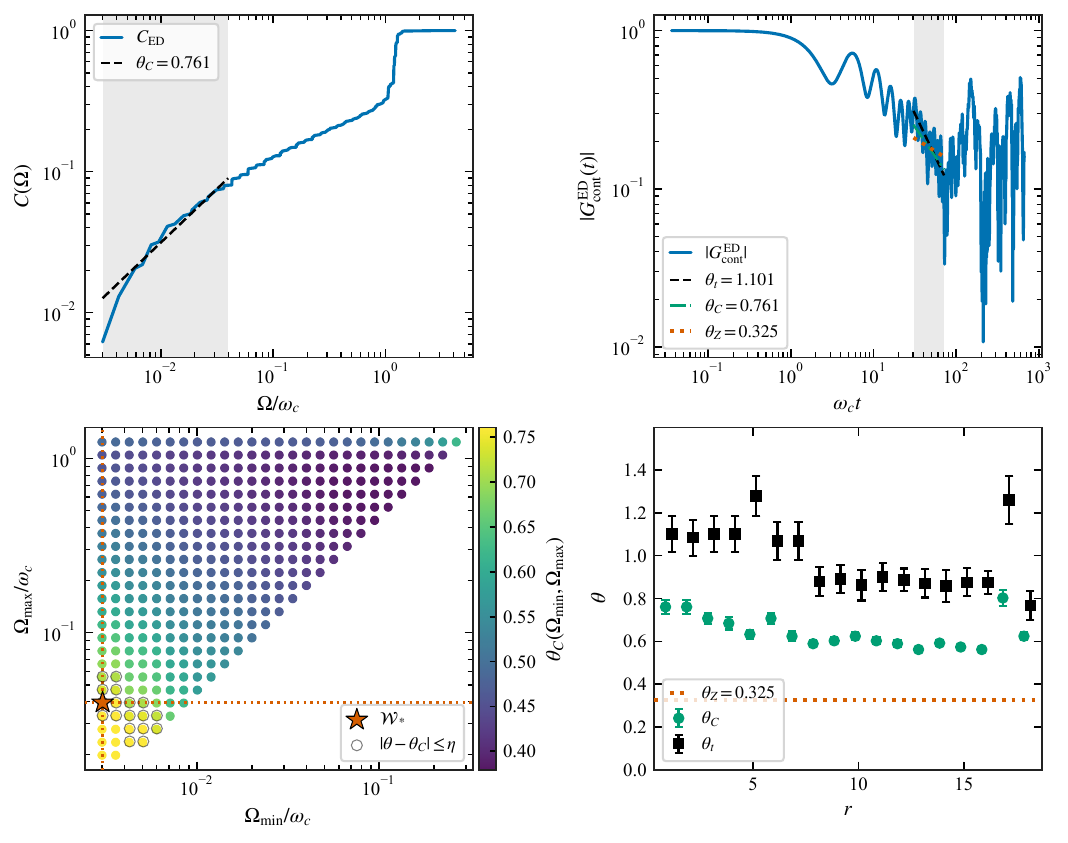}{Use this slot for the reciprocal time-domain and fit-window stability diagnostic figure.}
\put(-465,399){(a)}
\put(-196,399){(b)}
\put(-465,202){(c)}
\put(-196,202){(d)}
\caption{\textbf{Reciprocal time-domain diagnostic and fit-window stability of the continuum edge.}
(a) Energy-domain cumulative continuum \(C(\Omega)\) for a representative finite-tunnelling point. The dashed line shows the selected power-law fit \(C(\Omega)\sim\Omega^{\theta_C}\), and the shaded region marks the fitted energy window \(\mathcal W_*\).
(b) Continuum characteristic function \(|G_{\rm cont}^{\rm ED}(t)|\), computed directly from the same ED transition weights. The dashed line gives the best reciprocal-window time-domain fit, while the \(\theta_C\) and \(\theta_Z\) reference slopes are shown for comparison. This panel is diagnostic only because finite logarithmic spectra generate oscillations and recurrences.
(c) Fit-window stability map for the energy-domain exponent. Each point gives \(\theta_C(\Omega_{\min},\Omega_{\max})\) extracted from a different admissible fitting window. The star marks the selected window \(\mathcal W_*\), and open circles indicate windows whose exponent remains within the tolerance band around the selected value. The location of \(\mathcal W_*\) inside a region of comparable exponents shows that the continuum exponent is not obtained from a single cherry-picked interval.
(d) Reciprocal-window audit. The green points show \(\theta_C\) extracted from the ranked energy-domain windows, and the black points show the corresponding time-domain exponents \(\theta_t\) from reciprocal time windows. The dotted horizontal line marks the independently extracted elastic-overlap exponent \(\theta_Z\). The audit shows that the continuum and reciprocal-time diagnostics remain above the elastic diagnostic over the tested window set, while the energy-domain \(C(\Omega)\) extraction remains the primary exponent estimate.}
\label{fig:supp_time}
\end{figure*}

Figure~\ref{fig:supp_time} shows the reciprocal time-domain and fit-window stability checks for the continuum exponent.  Panel (a) shows the selected energy-domain cumulative-continuum fit and the fitted window. 
Panel (b) compares the corresponding reciprocal time-domain envelope with reference slopes set by \(\theta_C\) and \(\theta_Z\), emphasizing that the time-domain signal is used only as a diagnostic because finite logarithmic spectra contain oscillations and recurrences. 
Panel (c) maps the fitted continuum exponent over admissible fitting windows and verifies that the selected window is not isolated. 
Panel (d) compares the ranked energy-domain exponents with reciprocal-window time-domain exponents, showing that both diagnostics remain above the independently extracted elastic-overlap exponent over the tested window set.

\subsection*{I. Compact control summary: fixed-point anchor, infrared class, and ranked-window separation}

\begin{figure*}[t]
\centering
\includewithplaceholder{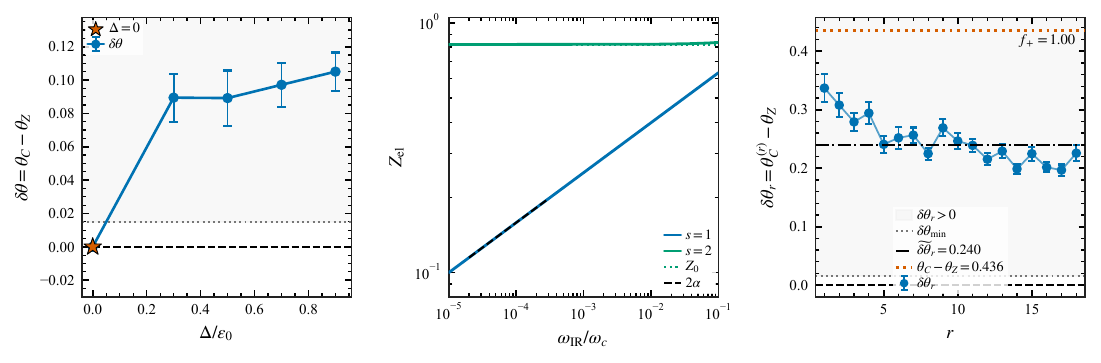}{Use this slot for the compact control-summary figure.}
\put(-469,164){(a)}
\put(-299,164){(b)}
\put(-130,164){(c)}
\caption{\textbf{Compact control summary for the finite-tunnelling edge.}
(a) Fixed-point anchor and finite-tunnelling splitting. At \(\Delta=0\), the exact independent-boson result gives \(\theta_C=\theta_Z=2\alpha\), hence \(\delta\theta=0\). For finite tunnelling, the extracted energy-domain separation becomes positive over the resolved scaling window.
(b) Infrared-class control. The Ohmic bath shows algebraic extinction of the elastic threshold weight, \(Z_{\rm el}\sim\omega_{\rm IR}^{2\alpha}\), while the super-Ohmic bath approaches a finite elastic residue \(Z_0\). This distinguishes the orthogonality edge from ordinary strong-coupling dressing.
(c) Ranked-window positive-separation audit. The plotted quantity is \(\delta\theta_r=\theta_C^{(r)}-\theta_Z\), where \(\theta_C^{(r)}\) is extracted from the \(r\)-th top-ranked admissible continuum fitting window. All shown ranked windows remain above the positive-separation threshold, giving \(f_+=1.00\). This supports the finite-window statement \(\theta_C>\theta_Z\) without using the time-domain diagnostic as the primary exponent evidence.}
\label{fig:supp_three_controls}
\end{figure*}

We include an additional compact control summary in Fig.~\ref{fig:supp_three_controls}. This figure gathers three checks that clarify the finite-tunnelling interpretation without introducing a new exponent extraction protocol.

First, the finite-tunnelling data are anchored at the exactly solvable independent-boson point. At \(\Delta=0\), the Ohmic fixed point gives
\begin{equation}
\theta_C=\theta_Z=2\alpha,
\qquad
\delta\theta=\theta_C-\theta_Z=0 .
\label{eq:supp_exact_anchor}
\end{equation}
The finite-\(\Delta\) data should therefore be interpreted as a splitting of two diagnostics that are locked together at the static boundary fixed point, rather than as an unrelated numerical trend.

Second, we compare the infrared behavior of the elastic threshold weight in Ohmic and super-Ohmic reservoirs. The Ohmic bath shows algebraic extinction,
\begin{equation}
Z_{\rm el}\sim \omega_{\rm IR}^{2\alpha},
\label{eq:supp_ohmic_control}
\end{equation}
whereas the super-Ohmic bath approaches a finite elastic residue \(Z_0\). This control separates the orthogonality-edge mechanism from ordinary strong-coupling dressing: the extinction of the elastic line is tied to the infrared class of the reservoir.

Third, we test the sign of the continuum--elastic separation across the best energy-domain fitting windows. Let \(\theta_C^{(r)}\) denote the continuum exponent extracted from the \(r\)-th top-ranked admissible fitting window, ordered by the same quality score used in the continuum analysis. We then define
\begin{equation}
\delta\theta_r=\theta_C^{(r)}-\theta_Z .
\label{eq:supp_ranked_delta}
\end{equation}
This ranked-window diagnostic is deliberately weaker, but safer, than requiring every window to reproduce the single selected value of \(\theta_C\). Its purpose is to check whether the positive separation from the elastic exponent persists across the leading admissible windows. In the displayed set, all ranked windows remain above the positive-separation threshold, supporting the finite-window statement \(\theta_C>\theta_Z\).

\end{document}